\documentclass{jfm}

\usepackage{graphicx}
\usepackage{natbib}

\ifCUPmtlplainloaded \else
  \checkfont{eurm10}
  \iffontfound
    \IfFileExists{upmath.sty}
      {\typeout{^^JFound AMS Euler Roman fonts on the system,
                   using the 'upmath' package.^^J}%
       \usepackage{upmath}}
      {\typeout{^^JFound AMS Euler Roman fonts on the system, but you
                   dont seem to have the}%
       \typeout{'upmath' package installed. JFM.cls can take advantage
                 of these fonts,^^Jif you use 'upmath' package.^^J}%
      }
  \else
  \fi
\fi

\ifCUPmtlplainloaded \else
  \checkfont{msam10}
  \iffontfound
    \IfFileExists{amssymb.sty}
      {\typeout{^^JFound AMS Symbol fonts on the system, using the
                'amssymb' package.^^J}%
       \usepackage{amssymb}%
         
       \let\ge=\geqslant  
      }{}
  \fi
\fi

\ifCUPmtlplainloaded \else
  \IfFileExists{amsbsy.sty}
    {\typeout{^^JFound the 'amsbsy' package on the system, using it.^^J}%
     \usepackage{amsbsy}}
    {}
\fi

\newsavebox{\astrutbox}
\sbox{\astrutbox}{\rule[-5pt]{0pt}{20pt}}

\title[Turbulence cascade]{What is turbulence, what is fossil turbulence, and which ways do they cascade?}

\author[C. H. Gibson]%
{C\ls A\ls R\ls L\ns H.\ns G\ls I\ls B\ls S\ls O\ls N$^1$%
  \thanks{Email address for correspondence: cgibson@ucsd.edu}}

\affiliation{$^1$Departments of MAE and SIO, Center for Astrophysics and Space Sciences, University of California at San Diego,
La Jolla, CA 92093-0411, USA\\}

\pubyear{2010}
\volume{650}
\pagerange{119--126}
\date{?; revised ?; accepted ?. - To be entered by editorial office}
\begin{document}

\maketitle

\begin{abstract}
Turbulence is  defined as an eddy-like state of fluid motion where the inertial-vortex forces  of the eddies are larger than any other forces that tend to damp the eddies out.  By this definition, turbulence always cascades from small scales (where the vorticity is created) to larger scales (where other forces dominate and the turbulence fossilizes).  Fossil turbulence is any perturbation in a hydrophysical field produced by turbulence that persists after the fluid is no longer turbulent at the scale of the perturbation.  Fossil turbulence patterns and fossil turbulence waves preserve and propagate information about previous turbulence to larger and smaller length scales.  Big bang fossil turbulence patterns are identified in anisotropies of temperature detected by space telescopes in the cosmic microwave background.  Direct numerical simulations of stratified shear flows and wakes show that turbulence and fossil turbulence interactions are recognizable and persistent.
\end{abstract}

\begin{keywords}
Authors should not enter keywords on the manuscript, as these must be chosen by the author during the online submission process and will then be added during the typesetting process (see http://journals.cambridge.org/data/\linebreak[3]relatedlink/jfm-\linebreak[3]keywords.pdf for the full list)
\end{keywords}

\section{Introduction}

Turbulence is notoriously difficult to define.  Most discussions of turbulence avoid any attempt at definition.  It is typically identified from lists of known symptoms, like a disease.  In this paper a narrow definition is proposed based on the inertial-vortex force $ \vec{v} \times \vec{w}$, where $\vec{v}$ is velocity and $\vec{w}$ is vorticity, \cite{gib96}.  Flows not dominated by inertial-vortex forces are non-turbulent, by this definition.  Persistent perturbations of vorticity, temperature, density etc. produced by turbulence at length scales no longer turbulent are termed fossil turbulence.  Because vorticity is produced at small viscous scales, turbulence always cascades from small scales to large (contrary to the usual assumption,\cite{tay38}).  Irrotational flows are therefore non-turbulent, even though irrotational flows typically provide the kinetic energy of the smaller scale turbulence.  Fossil turbulence and fossil turbulence waves preserve and propagate information about the original turbulence events to both larger and smaller scales and generally dominate mixing and diffusion processes in natural fluids and flows, \cite{gib10}.   By making a distinction between turbulence and fossil turbulence, Kolmogorovian universal similarity laws of turbulence and turbulent mixing are provided a physical basis.  

In the natural flows of the ocean, atmosphere, and cosmology, many examples of fossil turbulence are found.  Jet aircraft contrails are fossils of turbulence that persist hours after all turbulence from the airplane has ceased at length scales of the contrails.  Fossils of big bang turbulence created at $10^{-35}$ m Planck scales have persisted for $\sim 13.7$ billion years at length scales $\ge 10^{25}$ m, \cite{gib04, gib05}. Direct numerical simulations of stratified turbulent wakes have demonstrated fossil turbulence wave radiation and highly persistent perturbations at Ozmidov scales of the turbulence at fossilization, \cite{pham09, brucker10, diam05, diam11}.  Thousands of hydrodynamic phase diagrams with monitored oceanic conditions support the proposed description, \cite{leung11}, providing the basis of a generic stratified turbulence transport mechanism, \cite{gbkl11}.      
 
\section{Theory}
The conservation of momentum equations for collisional fluids may be written with the inertial-vortex force separated from the Bernoulli group of energy terms $B={v^2}/2 + p/{\rho + lw}$ and the various other forces of the flow.

\begin{equation}
  \partial{\vec{v}/\partial{t}}=-\nabla{B}+\vec{v} \times \vec{w}+\vec{{F_{viscous}}}+\vec{{F_{buoyancy}}}+\vec{{F_{Coriolis}}}+\vec{{F_{etc.}}}
  \label{Helm}
\end{equation}

For many natural flows the sum of the kinetic energy per unit mass ${v^2}/2$ and the stagnation enthalpy  $p/{\rho}$ along a streamline are nearly constant and the lost work $lw$ is negligible.  Turbulence is the class of fluid motions that arise when the first force term on the right $-\nabla{B} \sim 0$ and the the inertial vortex force $\vec{v} \times \vec{w}$ dominates all the other forces.

The ratio

\begin{equation}
Re = \vec{v} \times \vec{w} / \vec{{F_{viscous}}}
  \label{Helm}
\end{equation}

is the Reynolds number Re.  The best known criterion for the existence of turbulence is that $Re \ge Re_{crit}$, where $Re_{crit}$ is a universal critical value $\sim 10-100$ from the first Kolmogorov hypothesis. 

For turbulence to exist, inertial vortex forces must also overcome gravitational forces.  A Froude number $Fr$ ratio

\begin{equation}
Fr = \vec{v} \times \vec{w} / \vec{{F_{buoyancy}}}
  \label{Helm}
\end{equation}

 must exceed a universal critical value $Fr_{crit}$.  Many other dimensionless groups based on ratios of the inertial vortex force to other forces have been discussed.
 
We see that turbulence always must begin at the Kolmogorov length scale\cite{kol41} where the Reynolds number first exceeds a critical value and vorticity is produced by viscous forces.  The turbulence cascades to larger scales by vortex pairing until limited by one of the other fluid forces.  The Ozmidov scale at beginning of fossilization determines the maximum vertical scale of turbulence overturns and the scale of fossil turbulence waves radiated.  In self gravitational flows the buoyancy period is replaced by the gravitational free fall time. 

The proposed cascade of turbulent kinetic energy from small scales to large is the inverse of the  \cite{tay38} and \cite{lum92} cascades, both of which are physically backwards and misleading as seen by the growth of wakes, jets, and boundary layers.

\section{Observations}
\subsection{cosmology}

According to hydrogravitational dynamics HGD cosmology the universe began due to a turbulence instability at Planck conditions.  Evidence for a turbulent big bang event is emerging from space telescope observations of the cosmic microwave background CMB, \cite{gib10}.  The most recent evidence from WMAP CMB is in Figure 1, \cite{stark12}.

\begin{figure}
  \centerline{\includegraphics{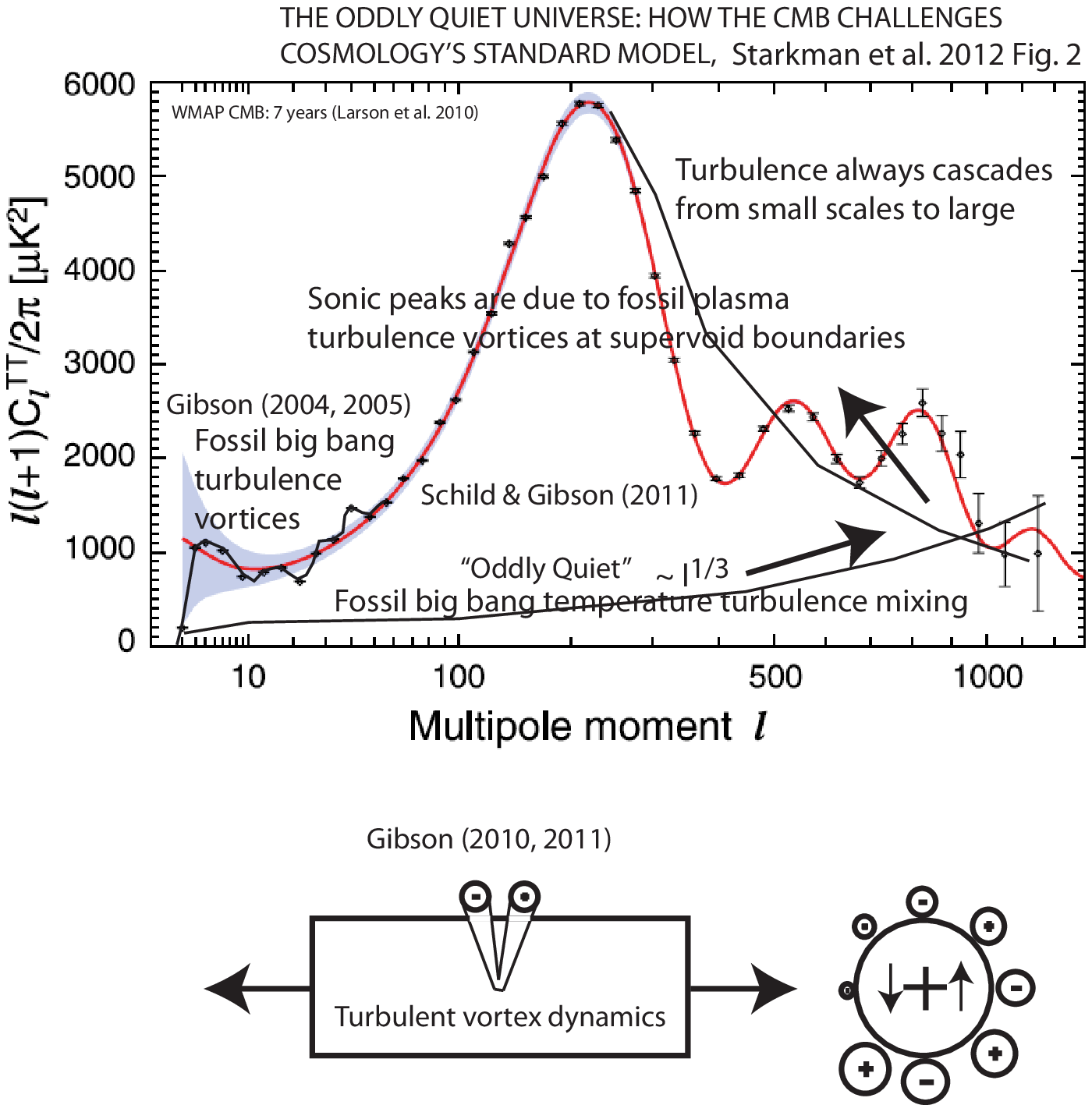}}
  \caption{Fossil turbulence in cosmic microwave background $CMB$ spectra, Starkman et al. 2012.}
\label{fig:ka}
\end{figure}

As shown in Fig. 1 (top), the measured spherical harmonics spectrum $C_\ell$ challenges the standard cosmological model, but is easy to understand as a fossil of big bang turbulence at multipole spatial frequencies $\ell$ near 10.    Multiple spectral peaks near $\ell = 10$ reflect vortex dynamics fossils of the big bang turbulence fireball, as shown in Fig. 1 (bottom), \cite{gib10,gsw11}. A similar fossil of turbulent supervoid fragmentation of the plasma along vortex lines explains $C_{\ell}$ for $\ell$ values larger than 200.  Baryon oscillations in cold dark matter potential wells are not needed to account for the sonic peaks observed at $\ell \ge 200$ and would be damped by the large photon viscosity of the plasma, even if cold dark matter potential wells were not mythical. 

Vortex line stretching provides the anti-gravitational negative stresses needed to extract mass-energy from the vacuum, as shown (arrows).  Dark energy is not needed.  Fossilization of big bang turbulence, \cite{gib05}, is due to phase changes of the Planck fluid as temperatures decrease from $10^{32}$ K to $\sim 10^{28}$ K, so that quarks and gluons, and large gluon viscosity, become possible, \cite{gib04}.  

Small scale turbulence during the plasma epoch was prevented by photon viscous forces from inelastic scattering of photons by electrons until the Schwarz viscous-gravitational scale $L_{SV}$ matched the scale of causal connection $L_H = ct$ at time $t=10^{12}$ s, where $c$ is the speed of light, \cite{gib96}.  Plasma fragmentation fossils persist as the largest objects, from superclusters to galaxies, \cite{gs10a, gs10b}.   The Corrsin-Obukhov fossil temperature turbulence mixing subrange $\ell(\ell + 1)C_{\ell} \sim \ell ^{1/3}$ at the bottom of Fig. 1 is a cascade from large scales to small (lower arrow) produced by big bang turbulent mixing.  The $sonic$ $peaks$ at $\ell \ge 200$ reflect the turbulence cascade (upper arrow) from small scales to large at the boundaries of gravitationally expanding super-cluster-voids.  The speed of expansion of voids is limited by the speed of sound $c/3^{1/2}$ in the plasma, and $\ell =200$ reflects the time  between first fragmentation time $t \sim 10^{12}$ s (30 000 years) and when the plasma becomes gas at $t \sim 10^{13}$ s.  

Because the kinematic viscosity $\nu$  decreases by a factor of $10^{13}$ from that of the plasma, the gas then fragments into planets in dense clumps that may become globular star clusters, \cite{gib96}.  This was inferred independently by \cite{schild96}  from his careful observations of gravitational lensing of a distant quasar by an intervening galaxy over a fifteen year period.  Because the twinkling period of quasar images is weeks (planets) rather than years (stars), Schild could infer that the mass of the galaxy is dominated by planets rather than stars.  Differences in brightness of the quasar images show the planets are in clumps.  
  
\subsection{Stratified shear flows and wakes}

As we have seen, the definitions of turbulence and fossil turbulence are important to understanding cosmology.  They are equally important to understanding terrestrial flows. High resolution microstructure detectors reveal most of oceanic mixing takes place in the fossil turbulence state, \cite{leung11}.  A generic mechanism of turbulent-fossil turbulence wave energy and information  transport from small scales to large is indicated, \cite{gbkl11}.  To capture the full stratified turbulence cascade mechanism, direct numerical simulations are required that resolve the Kolmogorov length scales where vorticity and turbulence are generated, \cite{diam05}, with a range of stratification periods ($\ge 1000$) sufficient to unequivocally demonstrate fossilization, \cite{diam11}.  

 \begin{figure}
  \centerline{\includegraphics{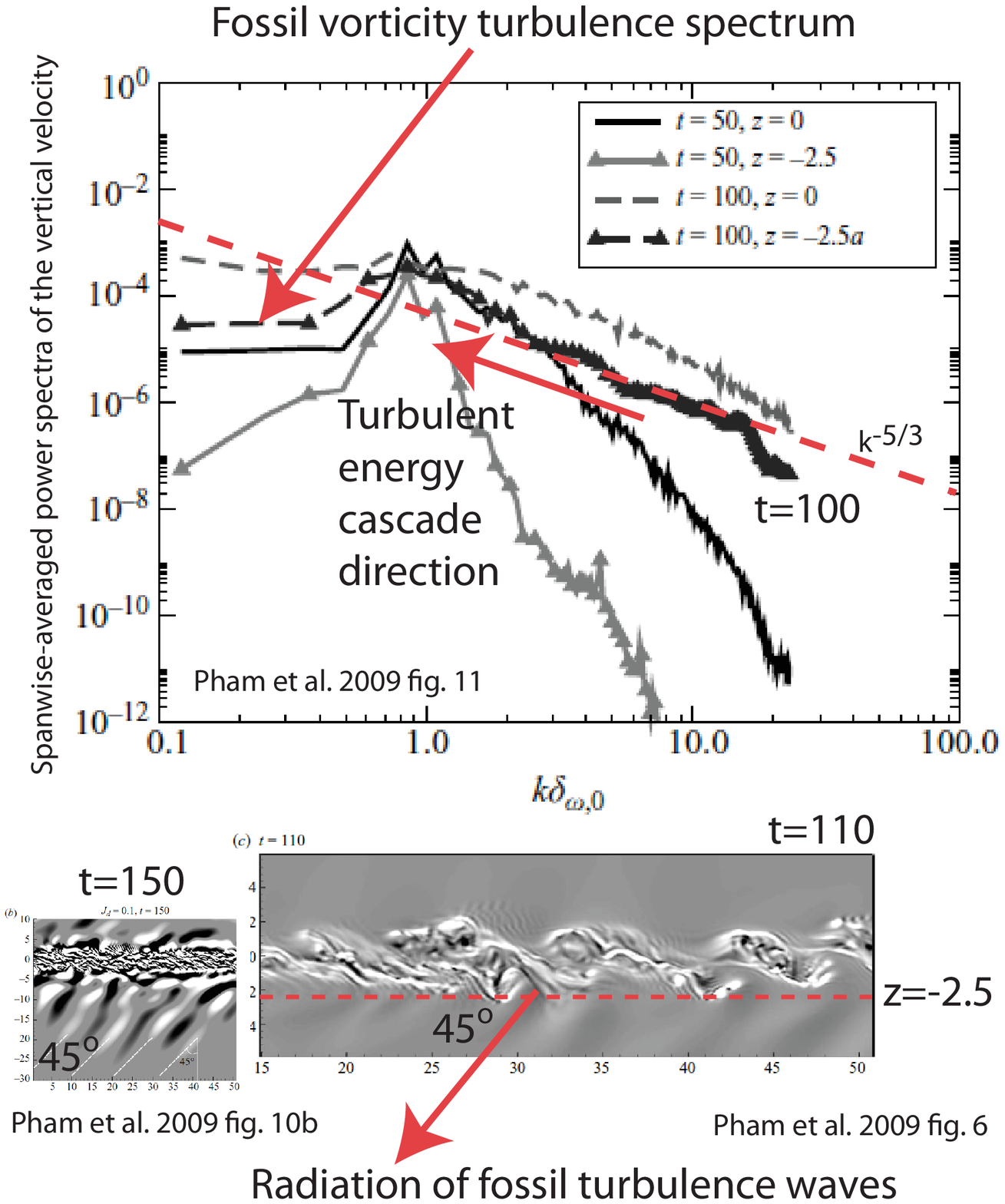}}
  \caption{Direct numerical simulation of $x,y,z,t$ turbulence and fossil turbulence waves formed by shear, \cite{pham09}.}
\label{fig:kd}
\end{figure}
 
 A fossil vorticity turbulence spectrum ($t=100, z= -2.5$) has been captured using direct numerical simulations of sheared stratified turbulence, as shown in Figure 2, from Fig. 11 in \cite{pham09}.  The dashed line with slope -5/3 suggests turbulence at small scales cascades to the Ozmidov scale at fossilization, the thickness of the shear layer.  Radiation of fossil turbulence waves at the Ozmidov scale and expected angle $45^o$ is shown at the bottom of Fig. 2.
 
Direct numerical simulations of stratified turbulent wakes are compared to non-stratified wakes by \cite{brucker10}, Figure 3.

\begin{figure}
  \centerline{\includegraphics{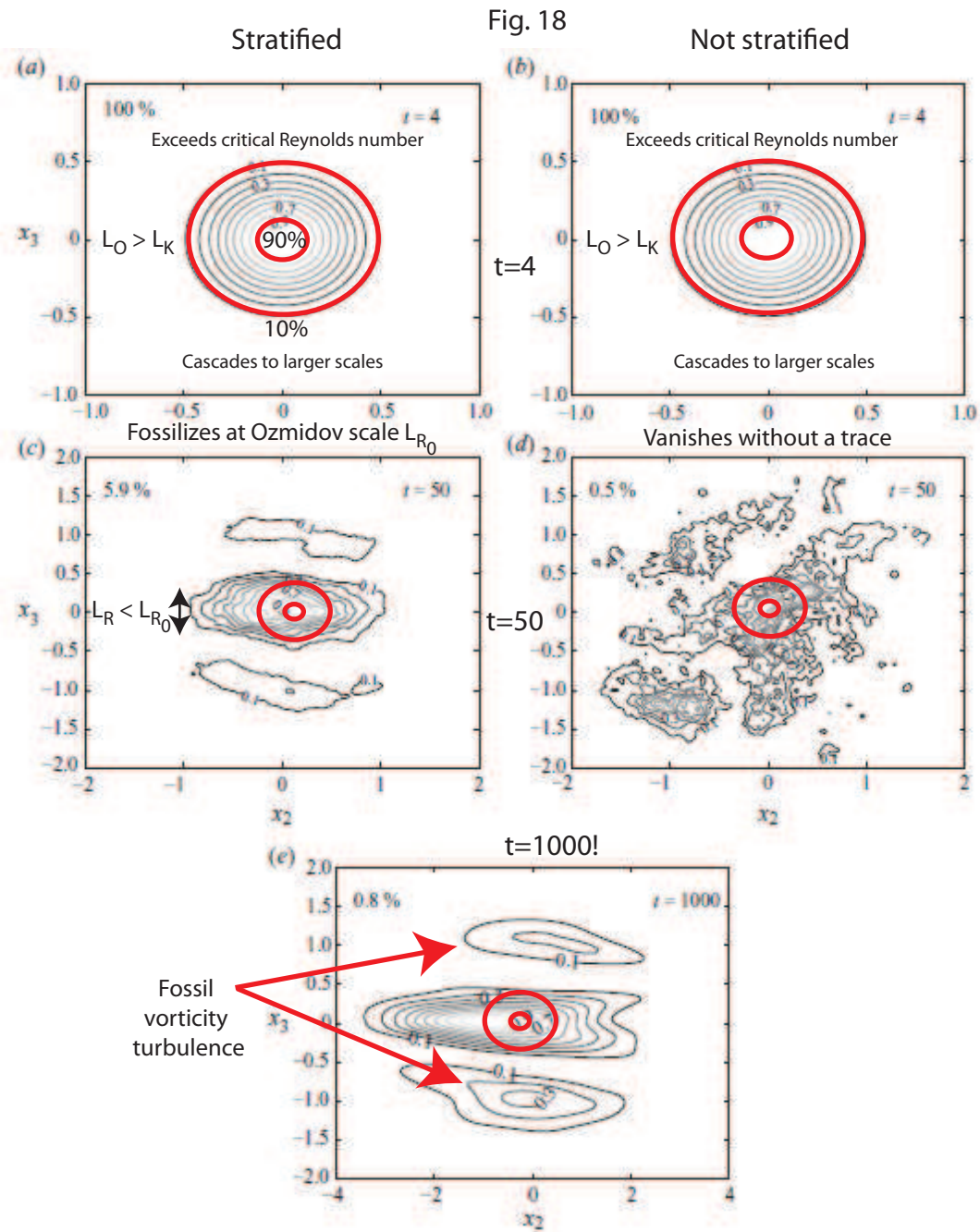}}
  \caption{Direct numerical simulation of $x,y,z,t$ turbulence and fossil turbulence waves formed by stratified and non-stratified wakes, Brucker et al. 2010.}
\label{fig:kd}
\end{figure}

As shown in Fig. 3 (top), turbulence forms in both stratified and non-stratified turbulent wakes when the Obukhov energy scale $L_O$ exceeds the Kolmogorov scale $L_K$.  The non-stratified turbulent wake cascades to larger length scales and vanishes without a trace.  The stratified turbulent wake fossilizes when the Ozmidov scale becomes smaller than the Ozmidov scale at beginning of fossilization.  Fossil turbulence waves propagate into surrounding stratified layers and persist for periods of time exceeding 1000 stratification periods (bottom).

Laboratory evidence demonstrating inverse turbulence cascade directions has existed for many years.  Figure 4 shows concentrated salt solution (arrow) injected into the turbulent mixing zone behind a cylinder in a water tunnel.  Refractive index fluctuations reveal the wake growth and mixing.  A micron scale single electrode conductivity probe (left) inserted along the tunnel test section axis detects salt concentrations.  It was found that small scale turbulent mixing begins immediately at small scales and continues downstream as a homogenizing volume bounded by a sharp interface surface (termed a superlayer by Stan Corrsin) separating the turbulent and non-turbulent fluid.  The superlayer is strongly distorted.  Unmixed fluid appears at least 5\% of the time.  Rare, large, conductivity spikes also detected at the centerline downstream position suggest fossil turbulence wave radiation has penetrated the turbulence superlayer, driven by the large baroclinic torques $\nabla \rho \times \nabla p / \rho ^2$ upstream.  The Taylor-Lumley cascades of turbulence and turbulent kinetic energy from large scales to small, \cite{tay38, lum92}, are falsified by these observations (and all others) showing the actual direction of these turbulence cascades is from small scales to large.   

\begin{figure}
  \centerline{\includegraphics{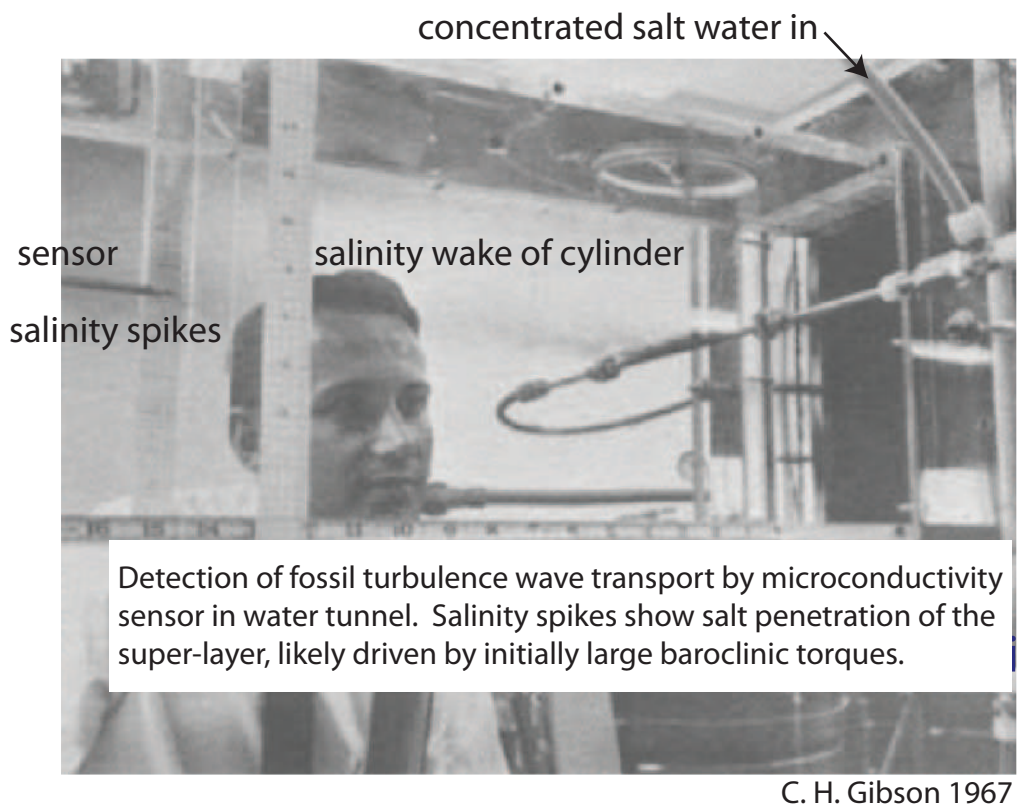}}
  \caption{Turbulent mixing by turbulence and fossil turbulence waves.}
\label{fig:ka}
\end{figure}

\section{Summary and Conclusions}

Turbulence must be defined in terms of the inertial vortex forces of its eddy motions for the universal similarity laws of Kolmogorov, Batchelor etc. to apply.  Turbulence always cascades from small Kolmogorov scales to larger Ozmidov etc. scales where other forces or fluid phase changes cause it to fossilize.  Fossil turbulence preserves information about previous turbulence and continues the mixing and diffusion started by turbulence.  Fossil turbulence waves dominate the transport of heat, mass, chemical species and information in natural fluids.  The 1938 G. I. Taylor concept that turbulence cascades from large scales to small is incorrect and misleading, and should be abandoned.  Irrotational flows in the Lumley spectral pipeline model should be identified as non-turbulent. 


\bibliographystyle{jfm}

\bibliography{jfm-instructions}

\end{document}